\documentclass[12pt,epsfig,amsmath,amssymb,pstricks]{article}
\usepackage{amsmath}
\usepackage{amsthm}
\usepackage{graphicx}
\usepackage{xcolor}
\usepackage{float}

\usepackage{hyperref}

\begin{document}

\title{\bf Metabolic limits on classical information processing by biological cells}

\author{{Chris Fields$^1$ and Michael Levin$^2$}\\ \\
{\it$^1$ 23 Rue des Lavandi\`{e}res}\\
{\it 11160 Caunes Minervois, FRANCE}\\
{fieldsres@gmail.com} \\
{ORCID: 0000-0002-4812-0744} \\ \\
{\it$^2$ Allen Discovery Center at Tufts University} \\
{\it Medford, MA 02155 USA} \\
{michael.levin@tufts.edu} \\
{ORCID: 0000-0001-7292-8084} 
}
\maketitle

\begin{abstract}
Biological information processing is generally assumed to be classical.  Measured cellular energy budgets of both prokaryotes and eukaryotes, however, fall orders of magnitude short of the power required to maintain classical states of protein conformation and localization at the \AA, fs scales predicted by single-molecule decoherence calculations and assumed by classical molecular dynamics models.  We suggest that decoherence is limited to the immediate surroundings of the cell membrane and of intercompartmental boundaries within the cell, and that bulk cellular biochemistry implements quantum information processing.  Detection of Bell-inequality violations in responses to perturbation of recently-separated sister cells would provide a sensitive test of this prediction.  If it is correct, modeling both intra- and intercellular communication requires quantum theory.
\end{abstract}


\section{Introduction}

Whether biological cells employ quantum coherence for information processing has been controversial since Schr\"{o}dinger \cite{schro:44} introduced the idea.  The question gained prominence when Hameroff and Penrose \cite{HP:96} proposed that neuronal microtubules function as quantum computers and Tegmark \cite{tegmark:00} countered that decoherence renders quantum computation in such systems infeasible; see \cite{hagan:02, mckem:09} for continuing discussion.  Recent studies in quantum biology have largely focused on the role of single-protein scale coherence in photoreception and magnetoreception in a variety of systems (see \cite{arndt:09, lambert:12, melkikh:15} for reviews).  Both experimental reproducibility and theoretical interpretation remain significant sources of controversy \cite{cao:20}.

Here we consider this question from the reverse direction: is it energetically feasible that all biological information processing is classical?  While some still contest it \cite{nijhout:90, longo:12}, since the pioneering efforts of Turing \cite{turing:52}, Polanyi \cite{polanyi:68}, and Rosen \cite{rosen:86} among others, the idea that biological processes are at bottom informational has become commonplace.  Such processes are generally considered both irreversible and classical, e.g. in analyses of variational free energy minimization underlying Bayesian optimization at multiple scales \cite{friston:13}.  Irreversibly encoding a classical bit has a free-energy cost of $\mathrm{ln2} k_B T$ \cite{landauer:61, bennett:82, landauer:99}, $k_B$ Boltzmann's constant and $T$ temperature.  A physical process $\psi_i \rightarrow \psi_j$ encodes one or more classical bits irreversibly whenever $\psi_i$ cannot be recovered from $\psi_j$.  Measurements of cellular energy budgets can, therefore, place upper limits on the numbers of bits that physical states of biological cells can irreversibly encode.  They thus provide a sensitive test of models based on classical information processing.

To investigate thermodynamic limits on classical encoding, we consider two models of cellular state for which supporting data are available: protein conformational state and protein localization state.  These states are obviously relevant to all cellular information processing, including signal transduction and gene expression as well as general metabolism; limiting models of cellular information processing to these degrees of freedom therefore provides a lower limit on the energetic cost of classical encoding.  We show that the energy budgets of both prokaryotic and eukaryotic cells are insufficient to classically encode bulk protein localization and conformation by factors of $10^{13}$ up to $10^{19}$.  We conclude that decoherence and hence classicality is restricted to low-dimensional components of the bulk cellular state space, and argue from previous results that these classical components are localized to either the cell membrane or to intercompartmental boundaries within the cell.  Boundaries, in other words, support classical communication, either within the cell or between the cell and its environment.  Away from such boundaries, bulk cellular information processing operates without classical encoding, i.e. via thermodynamically reversible, unitary processes that preserve quantum coherence.  We suggest that bulk coherence may be observable as entanglement between states of sister cells over macroscopic times following cell division.

\section{Decoherence sets the scale of classical encoding}

Fully-classical information processing at the molecular scale clearly requires an effectively-classical cellular state. Proteins, nucleic acids, polysaccarides, and even smaller organic molecules are standardly represented as having specific, effectively-classical conformations and locations, e.g. in the context of molecular dynamics calculations \cite{zwier:10, vlachakis:14}.  Molecular conformations and locations may evolve rapidly due to thermal noise at physiological temperature (for convenience we assume 310 K throughout) and may be unobservable except as samples from classical ensembles; however, this thermally-driven state change is represented as classical, thermodynamically irreversible mechanical motion, not as information-preserving unitary quantum evolution.  Both the characteristic classical dissipation ($\Delta t_{diss}$, ps to ns) and dynamic ($\Delta t_{dyn}$, 10s of ps to ms) timescales \cite{zwier:10, henzler:07} are, in particular, assumed to be significantly larger than the decoherence timescale $\Delta t_{dec}$ above which quantum coherence and hence thermodynamic reversibility is lost \cite{tegmark:00}.  

Standard models of decoherence, and hence the transition from quantum to classical behavior, support the assumption of classical molecular states by predicting decoherence time scales well below the scale of biologically significant processes, i.e. predict $\Delta t_{dec} << \Delta t_{dyn}$.  We can, for example, estimate the collisional decoherence timescale:

\begin{equation} \label{tdec-def}
\Delta t_{dec}(\Delta x) = \Gamma^{-1} (\Delta x)^{-2}
\end{equation}
\noindent
for center-of-mass position decohence at the scale $\Delta x$ using the long-wavelength limit for the scattering constant:

\begin{equation} \label{scattering}
\Gamma = (8/3) \hbar^{-2} (N/V) (2 \pi m)^{1/2} a^2 (k_B T)^{3/2}
\end{equation}
\noindent
given by \cite{schloss:07}, Eq. 3.73.  Here $N/V$ is the number density and $m$ the mass of scattering particles and $a$ is the radius of the scattering target.  For a typical protein with geometric volume of roughly 50,000 \AA$^3 = 5 \cdot 10^{-26}$ m$^3$ \cite{chen:15} in a water environment, $\Delta t_{dec}(\Delta x) \approx 6 \cdot 10^{-19}$ s for $\Delta x = 1$ nm, consistent with Tegmark's estimate, via an electrostatic-interaction model, of nm-scale position decoherence timescales for single ions in a cellular environment on the order of $10^{-20}$ s at 310 K \cite{tegmark:00}.  As molecular conformation is a function of the relative positions at Angstrom scales of amino-acid side chains or other small moities, individual conformation-angle decoherence times $\Delta t_{dec}(\Delta \varphi)$ for $\Delta \varphi \approx 10^{\circ}$ can be expected to be roughly four orders of magnitude longer, i.e. in the fs range.  For comparison, the time-energy uncertainty relation \cite{lloyd:00} gives the minimum dissipation time scale at $\Delta E_{th} = \mathrm{ln2} k_B T$ as $\Delta t_{diss} \geq \pi \hbar/ (2 \Delta E_{th}) \approx 50$ fs, roughly the time scale of molecular-bond vibrational modes and considerably shorter than dynamical timescales relevant to side-chain motion or peptide-bond hinging \cite{zwier:10}.

Decoherence effectively constructs discrete classical configuration spaces that macromolecules traverse in timesteps of $\Delta t_{dec}$.  Each amino-acid side chain, in particular, explores $\varphi$-angle space in discrete units of $\Delta \varphi$ in $\approx$ fs time steps as shown in Fig. 1.  Given $\Delta t_{dec}(\Delta x) << \Delta t_{dec}(\Delta \varphi)$, this variation in conformation angles is independent of the much faster (attosecond) exploration of center-of-mass position space.  Assuming independent peptide bonds, each individual angular evolution can be represented as a classical Markov process $\mathcal{M}_{ij}: \varphi_i \rightarrow \varphi_j$.  In this picture, measurements of conformational state, e.g. by binding to an enzymatic reaction center, sample a classical ensemble $\{ \varphi_k \}$ of temporal width $\Delta t_{meas}$ on the order of ns to $\mu$s, i.e. $\Delta t_{meas} >> \Delta t_{dec}$; averaging over this ensemble yields a coarse-grained, biologically-relevant outcome value $< \varphi >$.  Classical molecular dynamics calulations at \AA, fs scales model this Markovian evolution; the huge numbers of iterations needed to reach biologically-relevant time scales contribute to issues of both accuracy and feasibility \cite{zwier:10}.

\begin{figure}[H]
\centering
\includegraphics[width=14 cm]{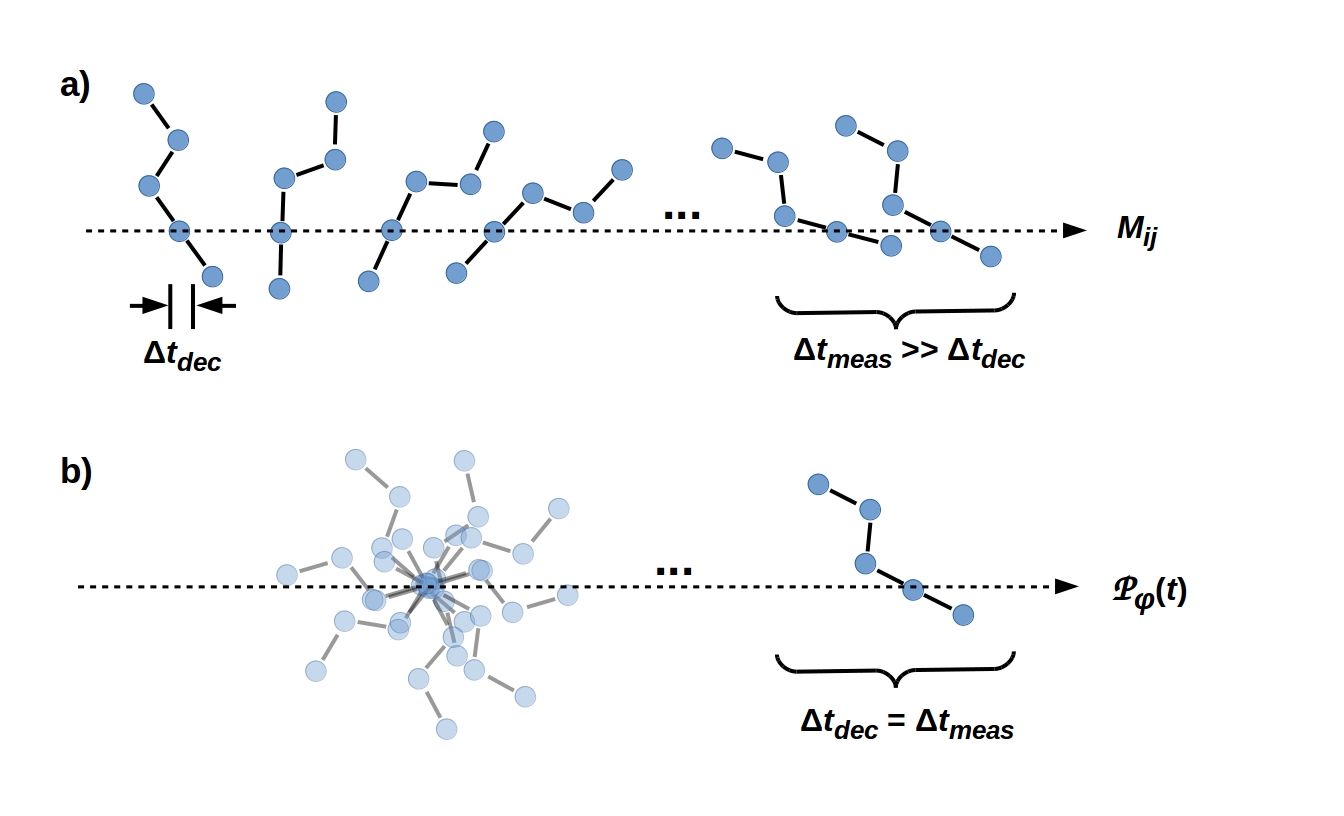}
\caption{a) Time evolution of a conformational state (ball-and-stick cartoon) driven by a classical Markov process $\mathcal{M}_{ij}$.  The conformational state is classical in every interval of length at least the decoherence time $\Delta t_{dec}$.  Measurements of the conformational state, e.g. by binding to an enzymatic reaction center, sample a classical ensemble $\{ \varphi_k \}$ of temporal width $\Delta t_{meas} >> \Delta t_{dec}$ to obtain a coarse-grained outcome $< \varphi >$.  b) Time evolution of a superposition $\varphi = \sum_i \alpha_i \varphi_i$ of molecular states $\varphi_i$ with amplitudes $\alpha_i$.  Binding to an enzymatic reaction center with characteristic time $t_{meas}$ effectively decoheres a particular outcome $\varphi_k$ with probability $|\alpha_k |^2$.}
\end{figure}

In the opposite, quantum limit of no decoherence, an isolated degree of freedom $\varphi$ evolves as a superposition $\varphi = \sum_i \alpha_i \varphi_i$ under the action of a unitary propagator $\mathcal{P}_{\varphi} (t)$.  In this case, binding to an enzymatic reaction center with characteristic time $\Delta t_{meas}$ introduces external interactions that effect decoherence, i.e. $\Delta t_{dec} =_{def} \Delta t_{meas}$, to some particular outcome $\varphi_k$ with probability $|\alpha_k |^2$.  Hybrid quantum mechanics/molecular mechanics (QM/MM) methods allow the treatment of small numbers of small molecules, e.g. water molecules and individual ions, as undergoing (approximately) unitary evolution within classical boundary conditions \cite{groenhof:13}.  The effective $\Delta t_{dec}$ is the fs-scale time step with which the boundary conditions, and hence interactions with the classical simulation are updated.

For the interior of a cell to occupy an effectively classical state at the macromolecular scale, all interior macromolecular degrees of freedom must be simultaneously decoherent.  Hence if $\rho_{cell}$ is the cellular state at the macromolecular scale, any decomposition $\rho_{cell} = \rho_{AB}$ into interior compartments $A$ and $B$ must be separable, i.e. $\rho_{AB} = \rho_A \rho_B$, and have entanglement entropy of zero.  The small-molecule (e.g. water) components of the intracellular environment must, in this case, absorb coherence and transport it to the extracellular environment efficiently enough that all collections $A$ and $B$ of macromolecular degrees of freedom mutually decohere within a timescale $\Delta t_{dec} (cell) << \Delta t_{dyn}$, the classical dynamic timescale.  It is not, however, clear that this is the case.  Straightforward models such as \eqref{scattering} assume independent collisions, an assumption that the presence of highly-structured hydration layers \cite{mattos:02, levy:06}, for example, renders unrealistic.  Such models also assume a very low density and infrequent interactions of scattering centers, i.e they assume $\Delta t_{dec} (cell) << \Delta t_{dyn}$ instead of demonstrating it.  Finally, standard decoherence models are both local and assume a single, monolithic environment; they do not provide an explicit mechanism for ``exhausting'' quantum coherence from the local decohering environment to an external environment located a macroscopic distance, many orders of magnitude larger than the decoherence length, away from the site of decoherence.

An explicit model of bulk decoherence of the molecular state of an entire cell is not feasible.  Our approach is instead to assume cellular-scale decoherence and hence effective classicality, and calculate the metabolic load required to maintain it.  We obtain orders-of-magnitude discrepancies even with minimal models of macromolecular state, indicating that bulk cellular decoherence is not, in fact, energetically feasible.

\section{Empirical data fix protein state-space dimensions}

The Boltzmann entropy, and hence Shannon information content \cite{shannon:48} of a state is determined not by the state itself, but by the number of possible states of the relevant degrees of freedom.  The classical information content of a bulk cellular state at the molecular scale is, therefore, determined by the number of possible molecular configurations, summed over all of the molecules in the cell.  As all known biochemical pathways involve both protein conformation and localization, these two classes of state variables provide a minimal model of macromolecular state.  Maintaining this state within a small, homeostatic region of the available state space -- in particular, preventing the denaturing of proteins into conformations incompatible with their normal functions -- is one of primary energetic tasks of the cell, and is the task that is prioritized as resources become limited \cite{fuente:11, kerr:19}.   Hence asking whether the cell can maintain a {\em classical} protein conformation and localization state at some given time scale is a criterial test: if it cannot, cellular information processing cannot be considered classical at that time scale.

We can, without loss of generality, treat state vectors as binary-valued and estimate their dimensionality $d$ in bits.  The dimensionality $d$(Conform) of the conformational component of the protein state space can then be estimated as the average number of bits required to specify a single protein conformation times the number of proteins.  Assuming an average length of 333 amino acids, specifying the conformation of a typical protein {\em ab initio} requires roughly $10^3$ bits \cite{fraenkel:93}; restricting each amino-acid side chain to one of three possible configurations would decrease this to an average of 10 bits/protein.  The protein content of cells can be estimated roughly as 40\% of dry mass \cite{ho:18}; measured values in representative species are reviewed in \cite{milo:13}.  Mean per-cell mass values for four taxonomic groups of prokaryotes \cite{makarieva:05} give estimated $d$(Conform) as shown in Table \ref{proks}.  

\begin{table}
\caption{Conformational state space dimension $d$(Conform) for four taxonomic groups of prokaryotes estimated using $10^3$ bits per protein \cite{fraenkel:93} and proteins/cell interpolated by cell volume from exemplars given in \cite{milo:13} Table 1.  Localization state space dimension $d$(Local) estimated using a typical protein geometric volume of 50,000 \AA$^3 = 5 \cdot 10^{-26}$ m$^3$ \cite{chen:15} and cell volumes estimated using 1 gm/ml from mass data given in \cite{makarieva:05} Table 1.  Mean power consumption from \cite{makarieva:05} Table 1 converted to bits/s using 1 bit $\equiv_{def} 3 \cdot 10^{-21}$ J.  Maximum computation rates $f_{max}$ are for fully-classical computation on the total protein state space.}
\label{proks}
\begin{center}
\begin{tabular}{| l | c | c | c | c |}
\hline
$~$  & Proteobacteria & Cyanobacteria & Firmicutes & Archaea \\
\hline
$d$(Conform) & 2.6 $\cdot 10^{9}$ & 4.2 $\cdot 10^{10}$ & 3.5 $\cdot 10^{8}$ & 3.5 $\cdot 10^{8}$ \\
\hline
$d$(Local) & 3.6 $\cdot 10^{7}$ & 5.6 $\cdot 10^{8}$ & 7.0 $\cdot 10^{6}$ & 6.0 $\cdot 10^{6}$ \\
\hline
$d$(Protein) & 9.4 $\cdot 10^{16}$ & 2.3 $\cdot 10^{19}$ & 2.4 $\cdot 10^{15}$ & 2.1 $\cdot 10^{15}$ \\
\hline
Power (fW) & 20 & 84 & 2.8 & 4.2 \\
\hline
Power (Mbits/s) & 6.7 & 28 & 0.93 & 1.4 \\
\hline
$f_{max}$ (Hz) & 7.1 $\cdot 10^{-11}$ & 1.2 $\cdot 10^{-12}$ & 3.9 $\cdot 10^{-10}$ & 6.7 $\cdot 10^{-10}$ \\
\hline
\end{tabular}
\end{center}
\end{table}

The dimensionality $d$(Local) of the localization component of the protein state space can be estimated from per-cell volume given an average protein volume of 50,000 \AA$^3$ \cite{chen:15} as employed above.  These estimates effectively assume that all proteins are cytosolic; as membrane-bound or other compartmentalized proteins must be actively transported to their functional locations, this is not an unreasonable assumption from a bioenergetic perspective.  Mean values for prokaryotes are given in Table \ref{proks}; $d$(Local) is roughly two orders of magnitude below $d$(Conform) across taxonomic groups.

The total protein state space dimension $d$(Protein) is computed assuming that conformation and localization are independent.  This is again an approximation, particularly for membrane-bound proteins.  As seen below, however, even decreasing values of $d$(Protein) by an order of magnitude to account for non-independence would have no qualitative effect.  

Tables \ref{euks} and \ref{humans} summarize values for $d$(Conform), $d$(Local), and $d$(Protein) for a representative unicellular eukaryotes \cite{fenchel:83} and for average adult human \cite{davies:13} and adult human cerebellar and cortical neurons \cite{herculano:11}, respectively.  Protein state spaces for eukaryotic cells are considerably higher than for prokaryotes, largely due to increased cell volume.  Values for neurons are for total neuronal complement within anatomical compartments, and hence average over very large Purkinje and canonical cortical neurons and much smaller granular cells, astrocytes, etc.

\begin{table}
\caption{State space dimensions for four representative unicellular eukaryotes calculated as in Table \ref{proks}, except cell volume data from \cite{fenchel:83}.  Mean power consumption from \cite{fenchel:83} Table 1 converted to bits/s using 1 bit $\equiv_{def} 3 \cdot 10^{-21}$ J.  Maximum classical computation rates $f_{max}$ calculated as in Table \ref{proks}.}
\label{euks}
\begin{center}
\begin{tabular}{| l | c | c | c | c |}
\hline
$~$  & {\em Ochromonas} & {\em Euglena} & {\em Bresslaua} & {\em Amoeba}  \\
\hline
$d$(Conform) & 1.5 $\cdot 10^{11}$ & 7.0 $\cdot 10^{12}$ & 3.3 $\cdot 10^{13}$ & $10^{15}$ \\
\hline
$d$(Local) & 5 $\cdot 10^{9}$ & 1.4 $\cdot 10^{11}$ & 6.6 $\cdot 10^{11}$ & 2.0 $\cdot 10^{13}$ \\
\hline
$d$(Protein) & 7.5 $\cdot 10^{20}$ & 9.8 $\cdot 10^{23}$ & 2.2 $\cdot 10^{25}$ & 2.0 $\cdot 10^{28}$ \\
\hline
Power (pW) & 22.5 & 210 & 1,650 & 10,000 \\
\hline
Power (Gbits/s) & 7.5 & 70 & 550 & 3,300 \\
\hline
$f_{max}$ (Hz) & $10^{-11}$ & 7.1 $\cdot 10^{-14}$ & 2.5 $\cdot 10^{-14}$ & 1.6 $\cdot 10^{-16}$ \\
\hline
\end{tabular}
\end{center}
\end{table}

\begin{table}
\caption{State space dimensions for average adult human cells from \cite{davies:13}, and adult human cerebellar and cortical neurons from \cite{herculano:11}.  Mean power consumption converted to bits/s using 1 bit $\equiv_{def} 3 \cdot 10^{-21}$ J; maximum classical computation rates $f_{max}$ calculated as in Table \ref{proks}.}
\label{humans}
\begin{center}
\begin{tabular}{| l | c | c | c |}
\hline
$~$  & Average Human & Cerebellar & Cortical \\
\hline
$d$(Conform) & 2 $\cdot 10^{12}$ & 2 $\cdot 10^{12}$ & 2 $\cdot 10^{12}$ \\
\hline
$d$(Local) & 8 $\cdot 10^{10}$ & 8 $\cdot 10^{10}$ & 8 $\cdot 10^{10}$ \\
\hline
$d$(Protein) & 1.6 $\cdot 10^{23}$ & 1.6 $\cdot 10^{23}$ & 1.6 $\cdot 10^{23}$ \\
\hline
Power (pW) & 1.5 & 42 & 750 \\
\hline
Power (Gbits/s) & 0.5 & 14 & 250 \\
\hline
$f_{max}$ (Hz) & 3.1 $\cdot 10^{-15}$ & 8.8 $\cdot 10^{-14}$ & 1.5 $\cdot 10^{-12}$ \\
\hline
\end{tabular}
\end{center}
\end{table}

Considering only protein conformation and localization clearly underestimates the dimensionality of the state space relevant to cellular information processing.  Highly localized concentration gradients of Ca$^{2+}$ and other ions, cofactors such as adenosine and guanine triphosphate (ATP and GTP), and other small molecules also play significant roles in regulating cellular state \cite{clapham:95, york:06, rosenbaum:09}, as does localized bioelectric state \cite{levin:12}.  Both the topological conformation and base-by-base inter-strand interaction states of nucleic acids are also significant, although typically on slower timescales \cite{kouzine:14, klemm:19}.  Hence using protein conformation and localization states to estimate the energetic cost of classical computation provides a lower limit only; more realistic values may be several orders of magnitude higher.

\section{Cellular energy budgets cannot support molecular-scale classical computation}

Classical computation requires updating a classical state, i.e. an $N$-bit string, on each computational clock cycle, independently of how many bits in the string change their values on each cycle.  Computation at a clock frequency $f$ incurs a minimal unit-time energetic cost (i.e. power consumption) of $N f \mathrm{ln2} k_B T$ \cite{landauer:61}.  Computing with 32 GB at a clock speed of 2.3 GHz at $T = 310$ K, for example, has a minimal cost of $\approx$ 1.7 W, roughly 80\% of the thermal design power of a commercial microprocessor with these specifications \cite{intel:18}.  For simplicity, we will abuse notation slightly to define ``1 bit'' of energy as $\mathrm{ln2} k_B T \approx 3 \cdot 10^{-21}$ J at $T = 310$ K, with a corresponding power unit of bits/$s$.

Tables \ref{proks}, \ref{euks}, and \ref{humans} summarize mean per-cell power consumption measurements for representative prokaryotes \cite{makarieva:05}, unicellular eukaryotes \cite{fenchel:83}, and average adult human cells \cite{davies:13} and adult human cerebellar and cortical neurons \cite{herculano:11}, respectively.  On average, eukaryotes consume three or more orders of magnitude more power per cell than prokaryotes, a difference attributable to both larger size and the availability of mitochondria specialized for respiration \cite{lane:10}.  Neurons, particularly cortical neurons, consume considerably more energy than average human cells, consistent with the brain's use of roughly 20\% of the whole-body energy budget in humans \cite{magistretti:15}. Significantly, and consistent with the above, this energy consumption is essentially independent of signalling activity, i.e. is primarily devoted to state maintenance \cite{hyder:13, bordone:20}.  

As tables \ref{proks}, \ref{euks}, and \ref{humans} indicate, cellular power consumption falls far short of that required to maintain a fully-classical, i.e. fully-decoherent state at the fs scale of molecular dynamics calculations or even the $\mu$s scale of protein domain motions.  Indeed the maximum classical computation rate $f_{max}$ achievable with the given power consumption falls short of 1 Hz by roughly $10^{10}$ for prokaryotes and small eukaryotic cells up to roughly $10^{15}$ for larger eukaryotic cells.  Even altogether ignoring protein localization, none of the cell types examined can support fully classical encoding of protein conformational state at 1 Hz.  We conclude that neither prokaryotes nor eukaryotes have the metabolic resources to support fully-classical information processing at the molecular scale.  Requiring classical computation rates in the kHz range of typical inter- and intra-cellular signaling processes restricts encoded classical states, and hence decoherence, to components spanning only $10^{-13}$ to $10^{-19}$ of the available protein state space.  The remaining components of the state space cannot undergo decoherence at kHz frequencies with the available metabolic resources.

\section{Classical encoding on intercompartmental boundaries}

The most straightforward interpretation of tables \ref{proks}, \ref{euks}, and \ref{humans} is that intracellular decoherence occurs not at the attosecond to fs time scales of molecular-scale fluctuations, but rather at the $\mu$s to ms time scales of intercompartmental or intercellular information exchange.  Assuming for simplicity that intercompartmental or intercellular information exchange in eukaryotes is mediated by transmembrane proteins, we can estimate the required classical-information encoding densities from transmembrane protein densities.  The density of tyrosine kinase receptors is roughly 550 per $\mu$m$^2$ on human HeLa cells and 1300 per $\mu$m$^2$ on CCRF-CEM cells \cite{chen:17}.  Assuming $\approx$ 10 pS conductance per channel \cite{negulyaev:00}, cortical neurons have between 5 (dendrites) and 250 (axon initial segment) Na$^+$ channels per $\mu$m$^2$ \cite{kole:08}.  Retinal rod cells can have on the order of 25,000 rhodopsin molecules per $\mu$m$^2$ in their layered outer-segment membranes \cite{molday:15}.  Hence it seems reasonable to estimate on the order of $10^5$ to $10^6$ transmembrane proteins on the outer cell membrane of a typical human cell, with comparable densities on other eukaryotic cells.  In the simplest model, each such protein encodes 1 bit per computational cycle.

Assuming spherical cells and power consumption from Tables \ref{proks}, \ref{euks}, and \ref{humans}, the maximum numbers of bits that can be processed at 1 kH by typical prokaryotic and eukaryotic cells are shown as a function of surface area in Fig. 2.  The data can be reproduced with a power law, bits(1 kH) = 100 + A$^{2.25}$, A the cell-surface area.  Hence while the range for human cells (green) is consistent with the estimates above, processing power grows at a rate faster than linear in either external surface area ($\sim$ A) or volume ($\sim$ A$^{1.5}$), suggesting that classical information is also encoded at high densities on intercompartmental boundaries.  

\begin{figure}[H]
\centering
\includegraphics[width=14 cm]{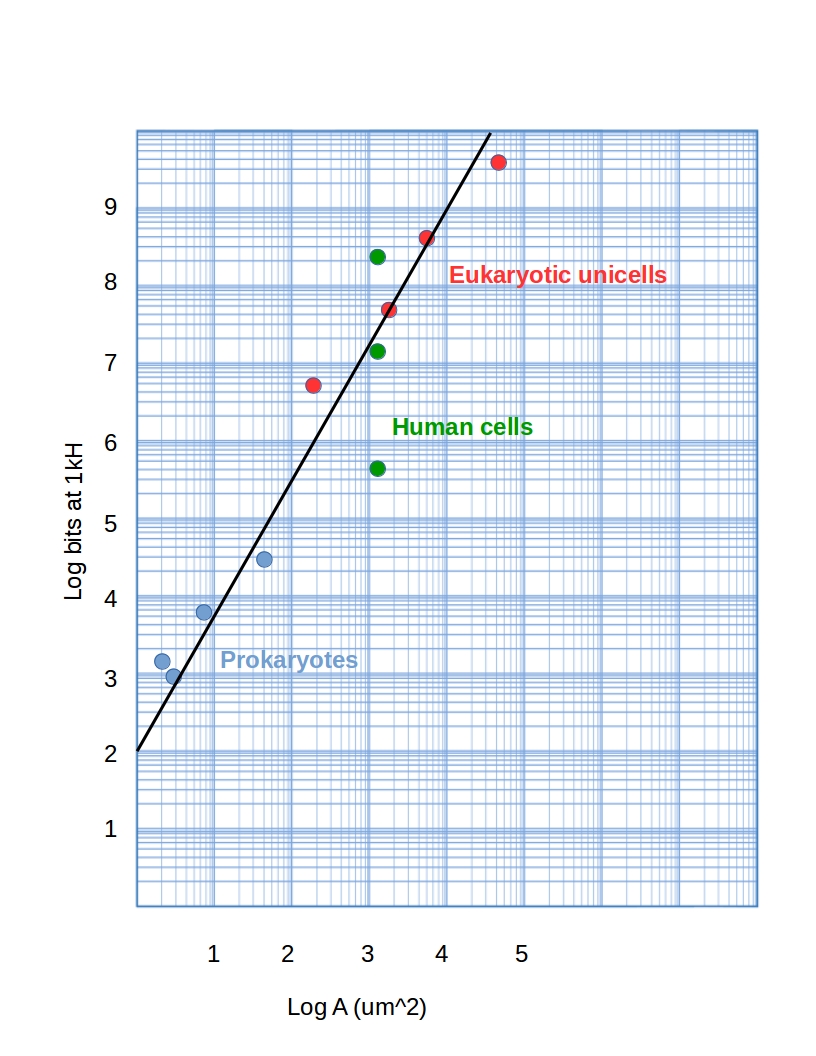}
\caption{Maximum bits processed at 1 kH as a function of surface area, assuming spherical cells and power consumption from Tables \ref{proks}, \ref{euks}, and \ref{humans}.  Solid line is the power law: bits(1 kH) = 100 + A$^{2.25}$, A cell-surface area, assuming a spherical cell.}
\end{figure}

Decoherence and hence classical encoding only at intercompartmental or intercellular boundaries is consistent with a model of bulk cellular compartments as weakly-interacting quantum systems occupying separable joint states.  Under these conditions, the inter-compartment interactions can be viewed as einselecting \cite{zurek:03, zurek:05} a computational basis and holographically encoding eigenvalues in this basis on the inter-compartment boundary \cite{fm:20, fgm:21}.  Labelling the compartments $A$ and $B$ as above, we can write the interaction as:

\begin{equation} \label{ham}
H_{AB} = \beta^k k_B T^k \sum_i^N \alpha^k_i M^k_i,
\end{equation}
where $k = ~A$ or $B$, the $M^k_i$ are $N$ Hermitian operators with eigenvalues in $\{ -1,1 \}$, the $\alpha^k_i \in [0,1]$ are such that $\sum^N_i \alpha^k_i = 1$, and $\beta^k \geq$ ln 2 is an inverse measure of $k$'s thermodynamic efficiency that depends on the internal dynamics $H_k$.  At each time step, $A$ and $B$ exchange $N$ bits of classical information specifying the current eigenvalue of $H_{AB}$, entirely independently of the bulk internal dynamics $H_A$ and $H_B$.  These $N$-bit encodings constitute the only decoherent, classical information in the system.  The energetic cost of computation is, in this case, only the energetic cost of classical encoding on the boundary; all information processing in each bulk compartment is unitary, implemented by the propagators $exp(-i/\hbar)H_k t$ and hence energetically free.

\section{Prediction: Entanglement between daughter cells}

A model in which decoherence is localized to intercompartmental boundaries suggests a strong and potentially testable prediction: that internal, bulk states of daughter cells may remain entangled for macroscopic times following cell division.  While long-lasting mirror symmetry of cytoskeletal components and hence motion patterns as well as cell-cycle correlation of sister cells has been observed \cite{albrecht:77} and numerous examples of epigenetic inheritance \cite{jablonka:92, cunliffe:15}, including epigenetic inheritance of organism-scale bioeletric state \cite{durant:17, durant:19} are now known, Bell-type experiments that directly test for state entanglement in recently-separated sister cells have not, to our knowledge, yet been performed.  If coherence is preserved as the above analysis indicates, perturbations of bulk biochemical state, e.g. targeting intermediate components of signal transduction pathways, in one daughter cell can be expected to affect the behavior of the other, biochemically and bioelectrically isolated, daughter cell.  In general, we may expect to observe correlations violating Bell-type inequalities \cite{bell:64} between distal structures or behaviors in living systems.  Such correlations may also be observable as Kochen-Specker contextuality \cite{kochen:67, mermin:93} or Leggett-Garg inequality \cite{emary:13} violations in time-series data.  Recent reports of Kochen-Specker contextuality, after correction for signalling, in human decision making \cite{cervantes:18, basieva:19} support the feasibility of detecting such effects with appropriately-designed experiments.

\section{Conclusion}

Consistent with single-molecule decoherence models, cellular biochemistry is standardly represented as classical at the \AA, fs scale of molecular dynamics calculations.  We have shown here that cellular energy budgets cannot support bulk classicality at this scale.  We suggest that biochemistry can only be treated as fully-classical at or near either the cell membrane or intercompartmental boundaries within the cell.  We predict on this basis that bulk-state entanglement may be observable between recently-separated sister cells.

While we have focused here on the measured energy budgets of extant cells, it is tempting to speculate that exceeding the limits imposed by classical thermodynamics is central, and perhaps definitional, to life as a phenomenon.  Searching for quantum effects in minimal, self-organizing biochemical systems \cite{engelhart:16, damiano:20} may shed light on this possibility.

\section*{Acknowledgements}
M.L. gratefully	acknowledges support by	the	Guy	Foundation.

\end{document}